\documentclass[12pt,epsffig]{article}
\usepackage[xdvi]{graphicx}
\newcommand{\beq}{\begin{equation}}
\newcommand{\eeq}{\end{equation}}
\newcommand{\bea}{\begin{eqnarray}}
\newcommand{\eea}{\end{eqnarray}}

\makeatletter
    
    \@addtoreset{equation}{section}
\makeatother

\begin{document}
\setlength{\baselineskip}{18pt}
\begin{titlepage}

\begin{flushright}
SAGA-HE-251 \\
\end{flushright}
\vspace*{3mm}
\begin{center}
{\bf\Large Meson-Nucleon Coupling from AdS/QCD \\ }
\vspace*{15mm}
{\large Nobuhito Maru\footnote{\tt maru@phys.chuo-u.ac.jp} and
Motoi Tachibana\footnote{\tt motoi@cc.saga-u.ac.jp}}

\vspace*{17mm}

{\em ${}^{*}$Department of Physics, Chuo University,
Tokyo 112-8551, Japan} \\
\vspace*{2mm}
{\em ${}^{\dagger}$Department of Physics, Saga University, 
Saga 840-8502, Japan} \\
\end{center}

\vspace*{30mm}

\begin{abstract}
In this manuscript, a unified approach to
hadron physics from holographic point of view is described. 
After introduction of a general setup for meson-nucleon
system based on the bottom-up approach of QCD (AdS/QCD),
as an illustration, we specifically examine meson-nucleon couplings. This is
an example of the notion we call ``holographic unification" in hadron physics.

\end{abstract}
\end{titlepage}

\section{Introduction}
Gauge/string duality conjectured by Maldacena \cite{Mal} is
one of the greatest developments in particle physics during the last decade.
It has not been proven mathematically yet, nevertheless has gained enough
credit through tons of nontrivial consistency checks \cite{review}.

The original version of the duality called AdS/CFT correspondence proposed
in \cite{Mal} is connecting type IIB supergravity theory compactified on
5 dimensional Anti-de Sitter space ($AdS_5$) and 5 dimensional sphere ($S^5$) to
 ${\cal N}=4$ supersymmetric (SUSY) Yang-Mills theory in 4 dimensions. Since
the beta function of ${\cal N}=4$ SUSY Yang-Mills is vanishing \cite{beta}, it is described
by conformal field theory (CFT). From this reason, the correspondence is
called AdS/CFT. After Maldacena's work, the original version of the AdS/CFT
duality has been extended to less supersymmetric cases
and even to non-supersymmetric ones. Besides the idea of the
AdS/CFT duality is deeply connected to that of ``holography", which
was proposed by several people \cite{holography}.

The most remarkable property of the  conjecture is
that a strongly-coupled quantum field theory can be described by a weakly-coupled
classical supergravity  and vice versa. This observation has immediately led
many people to study nonperturbative properties of field theories 
using supergravity descriptions. In nature, the most interesting
field theory whose nonperturbative dynamics is to be understood is 
Quantum Chromodynamics (QCD), i.e., dynamics of quarks and gluons.
It is empirically known in QCD that quarks, gluons and other color degrees of freedom
are confined into hadrons. Also chiral symmetry of QCD with massless quarks is 
spontaneously broken in vacuum and pions are the Nambu-Goldstone (NG)
bosons associated with it. Moreover at very high temperature/density,
QCD is expected to demonstrate some phase transitions into 
quark-gluon plasma (QGP)  \cite{qgp} 
and color superconducting (CSC) phases  \cite{csc}. These deconfined phases 
are intimately connected to physics of the early universe, heavy ion collisions
and compact stars. It is currently a major interest in QCD to comprehend 
space-time evolution of hadronic systems in those phases.

Shortly after Maldacena's work, many people including himself have tried to
apply the idea of AdS/CFT duality to QCD(-like)  theories.
For instance, Wilson loop operator, which is the order
parameter of confinement/deconfinement transition, in ${\cal N}$=4 SUSY Yang-Mills
was computed \cite{wilson} and a nontrivial dependence on the gauge coupling 
was found there. Besides the glueball mass spectra in QCD without matter
were calculated via AdS/CFT and compared with the results from Lattice QCD \cite{glueball}. 

Furthermore, as a crucial step toward more realistic QCD,
including quark degrees of freedom has been done by Karch and 
Katz \cite{karch-katz}. Shortly afterwards, the idea of \cite{karch-katz}
has been applied to problems, such as spontaneous
breaking of chiral symmetry (U(1) part), screening effect of quark-antiquark
potential via pair creation and meson mass spectra in different brane setups \cite{myers}.
Among them, the most successful one so far is ``Sakai-Sugimoto model" \cite{ss}.
In this model, QCD is realized by incorporating parallel $D8$-$\overline{D8}$ branes
perpendicular to $D4$ brane system\footnote{The D4 brane system itself has been originally introduced by Witten \cite{witten}}. 
Then one can obtain quarks with different chirality (left- and right-handed) from
open string excitations stretching between $D4$ and $D8$($\overline{D8}$) branes,
while gluons originate from $D4$-$D4$ strings.  In the dual gravity picture,  
$D4$ branes are replaced to a certain supergravity solution while $D8$ and $\overline{D8}$
branes have to be connected to each other (called probe approximation). 
This is the realization of chiral
symmetry breaking in geometric manner. Then the excitations from open strings
stretching $D8$ branes are interpreted as hadrons in QCD, such as pions, vector
and axial-vector mesons. Then vector mesons are the gauge bosons associated with
flavor symmetry and its spontaneous breaking provides vector meson masses
(5 dimensional Higgs mechanism). The model reproduces the experimental data
(masses, decay constants and various relations between physical parameters)
much better than expected. In addition, baryons can be treated as topological
excitations in this model \cite{ss-baryon}.

On the other hand, Sakai-Sugimoto model has surprising similarity with
deconstruction approach to QCD, proposed by Son and Stephanov \cite{deconstruction}.
In this approach, one starts from a set of 4 dimensional gauge theories with
bifundamental matter fields and take a ``continuum limit" according to
a certain prescription. Then one obtains 5 dimensional gauge theory
coupled to scalar field (dilaton) in some curved background.
This is called ``deconstruction of QCD".
In spite that the starting point is very different, the resultant effective action
gained from the procedure is essentially the same as that in Sakai-Sugimoto model.
The difference is that in the former the background metric is not known, but
in the latter it can be determined by solving the Einstein's equation.
In the deconstruction model, both vector and axial-vector mesons
appear as gauge bosons associated with flavor symmetry. This is
nothing but an extension of the idea of hidden local symmetry \cite{hls}
into an infinitely many vector bosons.

In such a situation, an interesting 5 dimensional model motivated
by AdS/CFT correspondence was proposed by Erlich et al. \cite{ekss} 
and Da Rold-Pomarol \cite{drp}. According to AdS/CFT dictionary,
there is a one-to-one correspondence between 4 dimensional operators 
in field theory side and 5 dimensional bulk fields in gravity theory side \cite{gkpw}.
Utilizing this property, one can start from 4 dimensional QCD 
and guess its 5 dimensional holographic dual \cite{shifman}. 
This approach is called ``AdS/QCD" or the bottom-up approach. 
Although there are an infinitely many
operators in QCD, as far as dynamics of chiral symmetry is concerned, 
it is enough to take into account several ones, such as left- and right-handed
flavor vector currents and chiral condensate. Vector current couples to a 
bulk vector field at the boundary while chiral condensate a bulk scalar field.
Then the resultant 5 dimensional action involves vector and axial-vector
mesons ($\rho, a_1$ etc) and scalar and pseudo-scalar mesons ($\pi, \sigma$ etc). 
There are several parameters in the model so that one can make some 
predictions after fixing those parameters from the experimental data.

After the works by \cite{ekss} and \cite{drp}, lots of applications and 
improvements of the original model have been accomplished. For
instance, deformation of bulk geometry \cite{bulk1} as well as bulk
Lagrangian \cite{bulk2}, incorporating other degrees of freedom into 
the original model such as scalar mesons \cite{scalar}, tensor mesons 
\cite{tensor}, glueballs \cite{glueballs}, baryons \cite{baryons} and 
other exotics \cite{exotics}, study of the Regge trajectries \cite{regge}, 
evaluation of inter-quark potential \cite{potential}, extension to finite 
temperature/density \cite{finite}, form factors \cite{ff}, phenomenological
studies \cite{pheno} and other general issues \cite{other}. 

The purpose of this manuscript is as follows: At first, we would like to
review the bottom-up approach (AdS/QCD) in detail with a simple setup
involving bulk meson and nucleon fields. Then we propose a scenario
of unification of hadron physics in higher dimensional space-time. There
all mesons and baryons are "unified" from holographic point
of view and finally it might be possible to  find out ``generalized" low-energy theorems,
where not only pions but also other vector and scalar mesons are
treated collectively at low energies.
Motivated by the scenario, we specifically try to investigate
general properties of meson-nucleon couplings and see
some useful relations. At the end, we summarize and give some
forthcoming perspectives.

\section{Review of bottom-up approach (AdS/QCD)}
\subsection{An example of  5D holographic model of QCD}

 As has been denoted in the previous section, according to
field-operator correspondence, there is one-to-one map
between 4 dimensional operators in field theory living on
the boundary and 5 dimensional bulk fields in gravity side.
In particular, a conserved current in 4 dimension is coupled
to a massless gauge boson on the boundary. Then the profile
of the field can be determined through the classical
equation of motion in the bulk with some appropriate
boundary conditions.

In the bottom-up approach of holographic QCD, the above
spirit is utilized by beginning from 4 dimensional QCD
and guessing its 5 dimensional holographic dual \cite{shifman}.
In QCD, there is an infinitely many operators, but as far as
chiral dynamics is concerned, we can restrict ourselves to
a certain set of operators, which are the left- and right-
conserved chiral current $J_{L, R}^{\mu a} \equiv \bar{q}_{L,R}\gamma^{\mu}t^a q_{L,R}$ ,
where $t^a$'s are $SU(N_f)$ flavor generators and chiral condensate $\bar{q}_R^{\alpha}q_L^{\beta}$,
where $\alpha$ and $\beta$ are $SU(N_f)$ flavor indices in fundamental representation.
With respect to those operators, one is able to assign 5 dimensional bulk fields.
This is summarized in Table below (Fig.1). 

\begin{center}
\begin{tabular}{ccccc} \hline \hline
4D:${\cal O}(x)$& 5D:$\phi(x, z)$ & $p$ & $\Delta$ & $(M_5)^2$  \\ \hline \hline
$\bar{q}_L\gamma^{\mu}t^{a}q_L$ & $A_{L\mu}^a$ & 1 & 3 & 0 \\
$\bar{q}_L\gamma^{\mu}t^{a}q_R$  & $A_{L\mu}^a$ & 1  & 3 & 0 \\
$\bar{q}_R^{\alpha}q_L^{\beta}$  & $(2/z)X^{\alpha\beta}$  & 0  & 3 & -3 \\
\hline\hline
\end{tabular}
\end{center}
\begin{center}
\textbf{Fig.1} Table of AdS/QCD (meson part) \cite{ekss, drp}
\end{center}

The system of our interest consists of $SU(N_f)_L \times SU(N_f)_R$ gauged flavor theory with a bulk scalar field 
$X$, which belongs to bifundamental representation under the gauge groups. 
The action we consider here as the meson sector is given by
\bea
S = \int d^5x \sqrt{-g} {\rm Tr} 
\left[
|DX|^2 -M_5^2 |X|^2 -\frac{1}{2g_5^2}(F_L^2 + F_R^2)
\right],
\eea
where the background metric is taken as 5 dimensional Anti-de Sitter space ($AdS_5$)
\footnote{This is not a unique choice, but just for brevity. Indeed various possibilities of deformed
bulk from $AdS_5$ have been proposed. For instance, see \cite{bulk1}.}:
\bea
ds^2 = \frac{1}{z^2} (\eta_{\mu\nu} dx^\mu dx^\nu - dz^2)
\qquad \qquad \varepsilon \le z \le z_m 
\eea
Here $\mu, \nu$ runs from 0 to 3 and $\varepsilon$ and $z_m$ are the inverse of UV and IR cutoff scales, respectively. 

Furthermore, a covariant derivative and a 5 dimensional bulk scalar mass are defined 
through the dictionary of AdS/CFT correspondence as follows:
\bea
&&D_M X = \partial_M X -i A_M^L X +i X A_M^R, \\
&&M_5^2 = \Delta (\Delta-4)=-3.
\eea
$\Delta$ denotes a conformal dimension of the operator in 4 dimension as seen in Fig. 1.
Then the classical equation of motion for $X$ is written as
\bea
\left[
-\frac{1}{z^3}\partial_\mu \partial^\mu + \partial_z \left( \frac{1}{z^3} \partial_z \right) -\frac{M_5^2}{z^5}
\right]X = 0,
\eea
which has a solution as follows:
\bea
X(z) = \frac{1}{2}(M z + \Sigma z^3),
\eea
where two integration constants, $M$ and $\Sigma$, are determined by the boundary conditions
at $z=\epsilon$ and $z=z_m$.
Here we have assumed that the classical solution of $X$ depends only on 
the fifth coordinate $z$ and used the relation $M_5^2=-3$ ($\Delta=3$) which is valid for scalar field. 
The physical meanings of $M$ and $\Sigma$ are the bare quark mass
(i.e., explicit breaking of chiral symmetry) and the chiral condensate (spontaneous chiral symmetry breaking). 
Below we take $M=m_q 1_{N_f\times N_f}$ and $\Sigma =\sigma 1_{N_f\times N_f}$ for simplicity.

\subsection{Gauge boson sector}

We next turn to the vector boson sector. 
For later use, it is convenient to rewrite the gauge bosons into the following base:
\bea
V_M = \frac{1}{2}(A^L_M + A^R_M), \quad 
A_M = \frac{1}{2}(A^L_M - A^R_M). 
\eea
To study the KK spectrum and the corresponding mode functions of the vector and axial vector mesons, 
it is enough to extract the quadratic terms in the action:
\bea
S_{{\rm quad}} &=& \int d^5x \frac{1}{z^5} \left[ 
-\frac{1}{2g_5^2}{\rm Tr}(\partial_M A_N^L - \partial_N A_M^L)^2 + (L \leftrightarrow R) 
\right]
\nonumber \\
&=& \int d^5x \left( - \frac{1}{4g_5^2 z} \right)
\left[
\partial_\mu V_\nu \partial^\mu V^{\nu} - \partial_\mu V_\nu \partial^\nu V^{\mu}
+ \partial_\mu A_\nu \partial^\mu A^{\nu} - \partial_\mu A_\nu \partial^\nu A^{\mu} 
\right. \nonumber \\
&& \left. 
- (\partial_\mu V_z - \partial_z V_\mu)^2 
- (\partial_\mu A_z - \partial_z A_\mu)^2
\right]. 
\eea

\subsubsection{Vector meson part}

Let us first  focus on the vector gauge boson sector. For that purpose,
we introduce the gauge-fixing term which cancels the mixing term proportional to $V_\mu V_5$: 
\bea
{\cal L}_{{\rm gf}}^V = - \frac{1}{2\xi_V g_5^2 z} \left[ \partial_\mu V^\mu -\xi_V z \partial_z \left(\frac{V_z}{z} \right) 
\right]^2
\eea
where $\xi_V$ is a gauge fixing parameter. 
Then we have the vector part of Lagrangian 
\bea
{\cal L}_V &=& -\frac{1}{4g_5^2z} 
\left[
\partial_\mu V_\nu \partial^\mu V^{\nu} - \partial_\mu V_\nu \partial^\nu V^{\mu}
- (\partial_\mu V_z - \partial_z V_\mu)^2
+\frac{1}{\xi_V}\left( \partial_\mu V^\mu -\xi_V z \partial_z \left(\frac{V_z}{z} \right) 
\right)^2
\right] \nonumber \\
&=& -\frac{1}{4g_5^2 z} 
\left[
\partial_\mu V_\nu \partial^\mu V^{\nu} - \partial_\mu V_\nu \partial^\nu V^{\mu}
-(\partial_\mu V_z)^2 - (\partial_z V_\mu)^2 
+ \frac{1}{\xi_V} (\partial_\mu V^\mu)^2 
+ \xi_V z^2 \left( \partial_z \left( \frac{V_z}{z} \right) \right)^2
\right] \nonumber \\
&=& -\frac{1}{4g_5^2z} V_\mu 
\left[
-\partial^2 \eta^{\mu\nu} + \partial^\mu \partial^\nu 
+ z \partial_z \left(\frac{1}{z} \right) \partial_z \eta^{\mu\nu} 
- 
\frac{1}{\xi_V} \partial^\mu \partial^\nu
\right]
V_\nu
\label{quadratic}
\eea
where the integration by parts are performed to arrive at the last expression. 
As a result, the boundary (surface) term should be vanished 
\bea
\left[ 
V_\nu \partial_z V^\nu
\right]_{\epsilon}^{z_m} = 0. 
\label{bdy}
\eea
Note also that $V_z$ decouples in a unitary gauge $\xi_V \to \infty$ because the mass of $V_z$ originated 
from the gauge-fixing term diverges (see, the last term in the second equation in (\ref{quadratic})),
\bea
m_{V_z}^2 \propto \xi_V z^2 \left( \partial_z \left( \frac{1}{z} \right) \right)^2 \to \infty. 
\eea
The equation of motion for $V_{\mu}$ in the unitary gauge can be easily read as
\bea
\left[ -\partial^2 \eta^{\mu\nu} + \partial^\mu \partial^\nu 
+ \eta^{\mu\nu} z \partial_z  \left( \frac{1}{z} \right) \partial_z \right] V_\nu = 0.
\label{bulkEOM}
\eea
Expanding in mode as $V_\nu(x,z) = \sum_n V_n^\nu(x) f_n^V(z)$ gives us the mode equation
\bea
\left[ m_n^2 + z \partial_z  \left( \frac{1}{z} \right) \partial_z \right] f_n^V(z) = 0 
\label{modeeom}
\eea
where $m_n$ denotes a four dimensional KK mass 
and a condition $\partial^\mu V_\mu=0$ is used since we focus on the transverse part of the gauge field.  

One obtains the mode function $f_n^V (z)$ as
\bea
f_n^V(z) = z \left[ c_1 J_1(m_n z) + c_2 Y_1(m_n z) \right] ,
\eea
where $c_{1,2}$ are integration constants  determined by  
the boundary conditions 
$f_n^V(\varepsilon) = \partial_z f_n^V(z_m) = 0$.
In more explicit form,
\bea
&&0=\varepsilon [ c_1 J_1(m_n \varepsilon) + c_2 Y_1(m_n \varepsilon) ], 
\label{bc1} \\
&&0=c_1 J_1(m_n z_m) + c_2 Y_1(m_n z_m) + z_m [ c_1 J_1'(m_n z_m) + c_2 Y_1'(m_n z_m) ]. 
\label{bc2}
\eea
Here $J_1$ and $Y_1$ are the Bessel functions.
These conditions can be converted to a condition for determining KK mass spectrum 
by eliminating $c_2$.
\bea
-\frac{1}{z_m} = \frac{J_1'(m_n z_m) Y_1(m_n \varepsilon) -J_1(m_n \varepsilon) Y_1'(m_n z_m)}
{J_1(m_n z_m) Y_1(m_n \varepsilon) -J_1(m_n \varepsilon) Y_1(m_n z_m)}. 
\label{KK}
\eea
To solve this equation, we make use of the following asymptotic forms of Bessel functions:
\bea
&&J_1(m_n \varepsilon) \simeq \frac{m_n \varepsilon}{2\Gamma(2)}, \quad 
Y_1(m_n \varepsilon) \simeq -\frac{2}{\pi m_n \varepsilon}
\eea
for $m_n \varepsilon \ll 1$ and 
\bea
&&J_1(m_n z_m) \simeq \sqrt{\frac{2}{\pi m_n z_m}} \cos \left(m_n z_m -\frac{3}{4}\pi \right), 
J_1'(m_n z_m) \simeq -\sqrt{\frac{2m_n}{\pi z_m}} \sin \left(m_n z_m -\frac{3}{4}\pi \right), \nonumber \\
&&Y_1(m_n z_m) \simeq \sqrt{\frac{2}{\pi m_n z_m}} \sin \left(m_n z_m -\frac{3}{4}\pi \right), 
Y_1'(m_n z_m) \simeq \sqrt{\frac{2m_n}{\pi z_m}} \cos \left(m_n z_m -\frac{3}{4}\pi \right), \nonumber \\
\eea
for $m_n z_m \gg 1$. 
Plugging these approximated expressions into (\ref{KK}) leads to
\bea
-\frac{1}{z_m} 
\simeq -m_n \tan \left(m_n z_m -\frac{3}{4} \pi \right). 
\label{KKmass}
\eea
The $n$-th KK mass is finally found  as
\bea
m_n \simeq \left(n - \frac{1}{4} \right) \pi z_m^{-1} \qquad (n=1,2,\cdots). 
\eea
As for the first excited state ($n=1$), namely $\rho$ meson, 
one obtains the mass as \cite{ekss,drp} 
\bea
m_\rho \simeq \frac{3}{4}\pi z_m^{-1} \simeq 2.4 z_m^{-1}. 
\eea
On the other hand, $\rho$ meson mode function is given by
\bea
f^\rho(z) = \frac{z J_1(m_\rho z)}{\sqrt{\int_\varepsilon^{z_m} dz z [J_1(m_\rho z)]^2}}. 
\eea

\subsubsection{Axial vector meson part}

Next let us turn to the axial vector sector. Since the longitudinal part of the
axial vector meson couples to the bulk scalar field $X$, 
one has to take into account
the quadratic part of Lagrangian involving both of them:
\bea
{\cal L}_{A+X} &=& -\frac{1}{4g_5^2 z} 
\left[
\partial_\mu A_\nu \partial^\mu A^\nu -\partial_\mu A_\nu \partial^\nu A^\mu 
-(\partial_\mu A_z -\partial_z A_\mu)^2
\right] \nonumber \\
&&+ \frac{v^2}{z^3}
\left[
(\partial_\mu P - A_\mu)^2 - (\partial_z P - A_z)^2
\right],
\label{A+X}
\eea
where a pseudoscalar field $P$ is a phase of the bulk scalar field $X$ defined as $X(x,z) = v(z) e^{iP}$
and $v(z)=\frac{1}{2}(m_q z+\sigma z^3)$. 

One also needs to add the gauge-fixing term such as
\bea
{\cal L}_{{\rm gf}}^A = -\frac{1}{2\xi_A g_5^2z}
\left[
\partial^\mu A_\mu -\xi_A z \partial_z \left(\frac{A_z}{z} \right) 
+ 2\sqrt{2} g_5^2 \frac{\xi_A}{z^2} v^2 P
\right]^2. 
\label{axialgf}
\eea
Thus, the resultant Lagrangian (quadratic part only) in a unitary gauge 
is given as
\bea
{\cal L}_{{\rm axial}} = -\frac{1}{4g_5^2z} 
A_\mu^a 
\left[
-g^{\mu\nu} \partial^2 + \partial^\mu \partial^\nu 
+ g^{\mu\nu} z \partial_z \left(\frac{1}{z} \right) \partial_z 
\right] A_\nu^a
+ \frac{v^2}{2z^3} (\partial_\mu P^a - A_\mu^a)^2
\eea
with vanishing boundary (surface) terms 
\bea
\left[ A_\mu \partial_z A^\mu \right]^{z_m}_\varepsilon = 0. 
\eea
The equation of motion is easily derived as 
\bea
\left[
- \frac{m_n^2}{z} - \partial_z \left(\frac{1}{z} \right) \partial_z 
+ \frac{2 g_5^2 v^2}{z^3}
\right]f_n^A = 0
\label{axialeom}
\eea
and the boundary conditions are given by
\bea
f_n^A(0) = \partial_z f_n^A(z_m) = 0. 
\label{bcaxial}
\eea
Note here that unlike the vector meson case, 
the axial meson sector depends on $v(z)$, i.e., the vacuum structure.
Plugging an explicit expression of $v(z)$ leads to
\bea
\left[
- \frac{m_n^2}{z} - \partial_z \left(\frac{1}{z} \right) \partial_z 
+ 2 g_5^2 \left( \frac{M}{2} + \frac{\sigma}{2} z^2 \right)^2
\right]f_n^A = 0. 
\eea
Obviously, the equation of motion for the axial vector has 
a $z$-dependent mass term and cannot be solved analytically. 
As an approximation, the bulk mass is supposed to be the brane localized mass at QCD brane ($z=z_m$). 
This is because $v(z)$ is localized towards the brane at $z=z_m$. 
In this approximation, the equation of motion in the bulk is the same as that of vector meson, 
but the boundary condition at $z=z_m$ is modified as follows. 
\bea
0 = \left. \left(\partial_z + 2 g_5^2 \frac{v^2}{z^2} \right) f_n^A(z) \right|_{z=z_m}.  
\label{modaxialbc}
\eea
We have already known a general solution in the bulk:
\bea
f_n^A(z) 
= c_1 z \left[J_1(m_n z) -\frac{J_1(m_n \varepsilon)}{Y_1(m_n \varepsilon)} Y_1(m_n z) \right] ,
\eea
where the boundary condition at $z=\varepsilon$ is imposed. 

By imposing the modified boundary condition (\ref{modaxialbc}) at $z=z_m$, 
we end up with finding the condition to determine the KK mass spectrum:
\bea
-\frac{1 + \frac{\sigma^2}{2} g_5^2 z_m^5}{z_m} = 
\frac{J_1'(m_n z_m) Y_1(m_n \varepsilon) - J_1(m_n \varepsilon) Y_1'(m_n z_m)}
{J_1(m_n z_m) Y_1(m_n \varepsilon) - J_1(m_n \varepsilon) Y_1(m_n z_m)}.
\label{axialmasscond} 
\eea
Likewise in the vector meson case, the right hand side of the condition can be approximated 
and we obtain the following result:
\bea
\tan \left(m_n z_m -\frac{3}{4} \pi \right) \simeq \frac{1 + \frac{\sigma^2}{2} g_5^2 z_m^5}{m_n z_m} 
\simeq \frac{g_5^2\sigma^2}{2m_n} z_m^4 
\eea 
where $z_m \gg 1$ is taken in the final approximation. 

Once if we make use of the mass of the first excited axial vector meson, $a_1$, 
\bea
m_{a_1} z_m \simeq 1230~{{\rm MeV}} \times \frac{2.4}{770~{{\rm MeV}}},
\eea
the value of the chiral condensate can be extracted from
\bea
(g_5 \sigma)^2 &\simeq& \frac{2m_{a_1}}{z_m^4} \tan \left(m_{a_1} z_m -\frac{3}{4} \pi \right),  
\eea
which leads to  
\bea
(g_5 \sigma)^{2/5} \simeq 775~{{\rm MeV}}. 
\eea
The normalized wave function of $a_1$ is given by
\bea
f_1^A(z) = \frac{zJ_1(m_{a_1}z)}{\sqrt{\int_\varepsilon^{z_m}dz z [J_1(m_{a_1}z)]^2}}. 
\label{modefcna1}
\eea

\subsection{Baryon sector with chiral symmetry breaking}

Now we turn to the baryon sector. 
Analysis of spin $\frac{1}{2}$ baryon in the context of holographic QCD has already been done,
for instance,  in \cite{HIY}. 
Here we follow their work and try to derive the meson coupling to nucleon. 
As we will show in the next section, this will give us an explicit example of
a unified approach from holographic viewpoint. 

Let us review the baryon sector in AdS/QCD \cite{HIY}. 
At first, the 5 dimensional bulk Lagrangian for baryon is provided as follows:
\bea
{\cal L}_{{\rm Baryon}} = \sqrt{-g} 
\left[
\frac{i}{2} \bar{N}_1 e_A^M \Gamma^A \nabla_M N_1
- \frac{i}{2} ( \nabla_M^\dag \bar{N}_1) e_A^M \Gamma^A N_1
-m_5 \bar{N}_1 N_1
\right],
\label{HIY}
\eea
where $\Gamma^A~(A=0,1,2,3,5)$ are 5D Dirac  matrices, 
$N_1$ is a Dirac fermion field transforming as $(N_f,1)$ under the gauge group $SU(N_f)_L \times SU(N_f)_R$, 
and the f\"unfvein is defined through $e_M^A = \frac{1}{z}\eta_M^A$ ($M$ is a space-time coordinate index 
and $A$ is a local Lorenz coordinate one). 
The covariant derivatives with respect to the general coordinate transformation and the gauge transformation are
given by 
\bea
\nabla_\mu &=& \partial_\mu + \frac{i}{4} \omega_\mu^{AB} \Gamma_{AB} -i A_\mu^L 
= \partial_\mu + \frac{1}{2z} \Gamma_z \Gamma_\mu -iA_\mu^L, \\
\nabla_z &=& \partial_z -i A_z^L 
\eea
where the nonvanishing spin connections are $\omega_M^{zA} = \frac{1}{z}\delta_M^A$. 
Note here that the bulk mass $m_5$ is related to the scaling dimension $\Delta$ of 
a corresponding boundary operator:
\bea
m_5^2 = \left( \Delta - 2 \right)^2. 
\label{bulkmass}
\eea

It should be emphasized that the above calculations do not care about effects of chiral symmetry breaking. 
Once taking into account the effects,  Yukawa interaction in the bulk should be introduced as follows:
\bea
{\cal L}_{{\rm Yukawa}} = -g_Y \bar{N}_1 X N_1 -g_Y \bar{N}_2 X^\dag N_2. 
\label{yukawa}
\eea
In this case, we should solve the following mode equations,
\bea
\left(
\begin{array}{cc}
\partial_z -\frac{\Delta}{z} & -\frac{g_Y v(z)}{z} \\
-\frac{g_Y v(z)^\dag}{z} &  \partial_z -\frac{4-\Delta}{z} \\
\end{array}
\right)
\left(
\begin{array}{c}
f_{1L} \\
f_{2L} \\
\end{array}
\right)
&=&-m_n
\left(
\begin{array}{c}
f_{1R} \\
f_{2R} \\
\end{array}
\right), \\
\left(
\begin{array}{cc}
\partial_z -\frac{4-\Delta}{z} & \frac{g_Y v(z)}{z} \\
\frac{g_Y v(z)^\dag}{z} &  \partial_z -\frac{\Delta}{z} \\
\end{array}
\right)
\left(
\begin{array}{c}
f_{1R} \\
f_{2R} \\
\end{array}
\right)
&=& m_n
\left(
\begin{array}{c}
f_{1L} \\
f_{2L} \\
\end{array}
\right),
\label{baryondirac}
\eea
where $f_{1L}$ is a wavefunction for $N_1$ satisfying the chirality
condition $i\Gamma^z N_{1L}=+N_{1L}$ and
similarly for others ($f_{1R}, f_{2L}$ and $f_{2R}$).
$m_n$ represents the KK masses for baryons. 

Here it might be instructive to show how the solution looks like even in the
case of no chiral symmetry breaking. 
In this case ($g_Y \rightarrow 0$), $f_1$ and $f_2$ sectors are completely decoupled to each other, 
then we end up with the following set of equations for $f_1$ sector:
\bea
&& \left[ \partial_z^2 -\frac{4}{z} \partial_z + \frac{6 + m_5 - m_5^2}{z^2} \right] f_{1L}^n(z) = - m_n^2 f_{1L}^n(z), 
\label{left} \\
&& \left[ \partial_z^2 -\frac{4}{z} \partial_z + \frac{6 - m_5 - m_5^2}{z^2} \right] f_{1R}^n(z) = - m_n^2 f_{1R}^n(z). 
\label{right}
\eea
The solution is obtained as
\bea
f_{1L(1R)}^n(z) = z^{5/2} \left[ c_1 J_{|m_5 \mp \frac{1}{2}|}(m_n z) + c_2 Y_{|m_5 \mp \frac{1}{2}|}(m_n z) \right]
\eea
where the right-handed mode function is obtained by replacing $m_5 \leftrightarrow -m_5$ 
in the left-handed mode function (the upper(lower) sign corresponds to the left(right)-handed mode). 

In our case, the scaling dimension of the baryon operator is $\Delta = \frac{9}{2}$, 
which implies that the bulk mass is fixed to be $m_5 = \pm \frac{5}{2}$. 
Following \cite{HIY}, the sign of the bulk mass is chosen such that 
the left(right)-handed zero mode comes from $N_1(N_2)$. 
We see from (\ref{bulkmass}) with $\Delta =\frac{9}{2}$ that $m_5 = \frac{5}{2} (-\frac{5}{2})$ for $N_1 (N_2)$. 
Therefore, the mode functions are found as
\bea
f_{1L}^n(z) &=& z^{5/2} \left[ c_{1L} J_2(m_n z) + c_{2L} Y_2(m_n z) \right], \\
f_{1R}^n(z) &=& z^{5/2} \left[ c_{1R} J_3(m_n z) + c_{2R} Y_3(m_n z) \right]. 
\eea
The boundary condition $f_{1L}^n(0)=0$ tells us 
\bea
c_{2L}^n = -\frac{J_2(m_n \varepsilon)}{Y_2(m_n \varepsilon)} c_{1L}^n 
= \frac{\pi}{2} \left( \frac{m_n \varepsilon}{2} \right)^4 c_{1L}^n \to 0~(\varepsilon \to 0) 
\eea
where the asymptotic form of Bessel function are used
\bea
J_2(m_n \varepsilon) \simeq \frac{1}{\Gamma(3)} \left(\frac{m_n \varepsilon}{2} \right)^2, \quad 
Y_2(m_n \varepsilon) \simeq -\frac{1}{\pi} \left(\frac{2}{m_n \varepsilon} \right)^2. 
\eea
Thus, we get
\bea
f_{1L}^n(z) = c_{1L}^n z^{5/2} J_2(m_n z). 
\eea
The right-handed mode function can be obtained from (\ref{baryondirac})
\bea
f_{1R}^n(z) = -\frac{1}{m_n} \left( \partial_5 -\frac{2+\frac{5}{2}}{z} \right)(c_{1L}^n z^{5/2} J_2(m_n z)) 
= c_{1L}^n z^{5/2} J_3(m_n z). 
\eea
Another boundary condition $f_{1R}^n(z_m)=0$ leads to the condition for determining the KK mass spectrum, 
\bea
J_3(m_n z_m) = 0. 
\eea

Similar analysis can be also applied to $N_2$ by replacing $m_5 = -5/2$ and $L \leftrightarrow R$. 
The boundary condition $f_{1R}^n(0)=0$ says
\bea
c_{2R}^n = -\frac{J_3(m_n \varepsilon)}{Y_3(m_n \varepsilon)} c_{1R}^n 
= -\frac{\pi}{\Gamma(4)} \left(\frac{m_n \varepsilon}{2} \right)^6 c_{1R}^n \to 0~(\varepsilon \to 0). 
\eea
Thus, we find 
\bea
f_{2R}^n(z) = c_{1R}^n z^{5/2} J_2(m_n z). 
\eea
The left-handed mode function is also obtained as
\bea
f_{2L}^n(z) = \frac{1}{m_n} \left( \partial_5 -\frac{2+\frac{5}{2}}{z} \right)(c_{1R}^n z^{5/2} J_2(m_n z)) 
= -c_{1R}^n z^{5/2} J_3(m_n z). 
\eea
The KK mass spectrum is obtained from another boundary condition $f_{2L}^n(z)=0$, 
\bea
J_3(m_n z_m) = 0.
\eea
In summary, we have obtained the following wave functions and the KK mass spectrum of the baryon. 
\bea
&& f_{1L}^n(z) = c_1^n z^{5/2} J_2(m_n z), \quad f_{1R}^n(z) = c_1^n z^{5/2} J_3(m_n z), \\
&& f_{2L}^n(z) = -c_2^n z^{5/2} J_3(m_n z), \quad f_{2R}^n(z) = c_2^n z^{5/2} J_2(m_n z), 
\eea
and 
\bea
J_3(m_n z_m) \propto \cos \left(m_n z_m -\frac{7}{4} \pi \right) = 0 
\to m_n \simeq \left( n - \frac{3}{4} \right) \pi z_m^{-1}.  
\eea
The normalization constants $c_{1,2}$ are fixed as follows. 
\bea
\int_0^{z_m} dz z |J_3(m_n z)|^2 &=& 
\left[
\frac{z^2}{2} \left\{ \left( 1-\frac{9}{(m_n z)^2} \right) J_3(m_n z)^2 
+ \underbrace{(J_3'(m_n z))^2}_{\left(J_2(m_n z) -\frac{3}{m_n z}J_3(m_n z)\right)^2} \right\}
\right]_0^{z_m} \nonumber \\
&=& \frac{z_m^2}{2} J_2^2(m_n z_m), \\
\int_0^{z_m} dz z |J_2(m_n z)|^2 &=& 
\left[
\frac{z^2}{2} \left\{ \left( 1-\frac{4}{(m_n z)^2} \right) J_2(m_n z)^2 
+ \underbrace{(J_2'(m_n z))^2}_{\left(-J_3(m_n z) + \frac{2}{m_n z}J_2(m_n z) \right)^2} \right\}
\right]_0^{z_m} \nonumber \\
&=& \frac{z_m^2}{2} J_2^2(m_n z_m)
\eea
where we note $J_{2,3}(0) = J_3(m_n z_m)=0$. 
Thus, the normalization constant is found as
\bea
|c_{1,2}| = \frac{\sqrt{2}}{z_m J_2(m_n z_m)}. 
\eea
In the case with chiral symmetry breaking, 
$N_1$ and $N_2$ are coupled with each other. 
Then one cannot solve the equations of motion analytically.
We have to perform numerical analysis.

\section{Toward a unified description of hadron physics from holographic QCD}
\subsection{``Holographic unification" of hadrons}

As we have already seen in the previous sections, the idea of gauge/string
duality, where one of the concrete realizations is the AdS/CFT correspondence,
naturally leads us to the conjectured scenario of an unified description of hadrons in
holographic way. According to the scenario, all the mesons and baryons (and
even some other exotics) are ``unified" in 5 dimensional curved space-time.
So that various physical quantities in 4 dimensions, which seem to be
different mutually,  might be understood collectively. We shall call this
``holographic unification" of hadrons. It means not only the unification
of matter in the sense of Kaluza-Klein, but also the unification of 
their physical properties such as couplings, low-energy behaviors and so on.

As such an example, we consider below 
the meson-nucleon couplings from unified point of view.

\subsection{Meson-nucleon couplings from unified point of view}

The basic spirit of AdS/QCD is that one writes down all the possible terms
allowed by symmetries in the bulk and evaluates different physical
quantities defined on the boundary. Note here that since we are dealing with higher
dimensional field theory, renormalizability cannot be a guiding principle any more.

Now let us consider the meson-nucleon couplings based on the system
argued in the previous section.
First, one starts with the pion-nucleon coupling. In terms of holographic QCD, it is derived from Yukawa
coupling (\ref{yukawa}) as well as  the 5th component of the covariant
derivative (\ref{HIY}):
\bea
{\cal L}^{\pi NN} &=& \int_0^{z_m} dz \sqrt{-g} 
\left[
\frac{i}{2} \bar{N}_1  \Gamma^5 (-iA_z^L) N_1 
- \frac{i}{2} (iA_z^L \bar{N}_1)  \Gamma^z N_1 + (L \leftrightarrow R, \ 1 \leftrightarrow 2)
\right] \nonumber \\
&& +\int_0^{z_m} dz \sqrt{-g} 
\left[
-g_Y \bar{N}_1  X N_1 
- g_Y \bar{N}_2  X^{\dagger} N_2
\right]. 
\eea
From this, one can read the pion-nucleon coupling as  \cite{HIY}
\bea
g_{\pi N^l N^l} = \int_0^{z_m} dz \frac{1}{z^4} 
\left[ f^{\pi}(f_{1L}^{l*}f_{1R}^{l}-f_{2L}^{l*}f_{2R}^{l}) 
-\frac{g_Y}{2v(z)zg_5^2}\partial_z
\left ( \frac{f^{\pi}}{z} \right )(f_{1L}^{l*}f_{2R}^{l}-f_{2L}^{l*}f_{1R}^{l}) \right] 
\label{pinn}
\eea
where $f^\pi$ is a mode function of pion originated from the 5th component of the axial vector gauge boson $A_z$. 

On the other hand, the vector and axial-vector meson-nucleon couplings in holographic QCD
are supplied in two ways. The first one originates
from the covariant derivative of the gauge interactions:
\bea
{\cal L}^{{\rm gauge}} &=& \int_0^{z_m} dz \sqrt{-g} 
\left[
\frac{i}{2} \bar{N}_1 e_A^M \Gamma^A (-iA_M^L) N_1 
- \frac{i}{2} (iA_M^L \bar{N}_1) e_A^M \Gamma^A N_1 \right. \nonumber \\
&& \left. + \frac{i}{2} \bar{N}_2 e_A^M \Gamma^A (-iA_M^R) N_2 
- \frac{i}{2} (iA_M^R \bar{N}_2) e_A^M \Gamma^A N_2
\right] \nonumber \\
&\supset&
\int_0^{z_m} dz \frac{1}{z^4} 
\left[
\bar{N}_1 \gamma^\mu V_\mu N_1
+ \bar{N}_2 \gamma^\mu V_\mu N_2
+\bar{N}_1 \gamma^\mu A_\mu N_1
- \bar{N}_2 \gamma^\mu A_\mu N_2
\right] \nonumber \\
&=& 
\int_0^{z_m} dz \frac{1}{z^4} f_n^V 
\left( |f_{1L}^l|^2 + |f_{1R}^l|^2 \right) \left[ \bar{N}_1^l \gamma^\mu V_\mu^{n} N_1^l 
+ \bar{N}_2^l \gamma^\mu V_\mu^{n} N_2^l
\right] \nonumber \\
&&+ \int_0^{z_m} dz \frac{1}{z^4} f_n^A 
\left( |f_{1L}^l|^2 + |f_{1R}^l|^2 \right) \left[ \bar{N}_1^l \gamma^\mu A_\mu^{n} N_1^l 
- \bar{N}_2^l \gamma^\mu A_\mu^{n} N_2^l
\right] ,
\eea
where the parity properties of the mode functions $f_{1L}=f_{2R}, f_{1R}=-f_{2L}$ are used \cite{HIY, HKSY}. 

In addition, the vector and axial-vector meson-nucleon couplings are given by the Pauli term:
\bea
{\cal L}^{{\rm Pauli}} &=& c \int_0^{z_m} dz \sqrt{-g} 
i \left[
\bar{N}_1 \Gamma^{MN} F_{MN}^L N_1 - \bar{N}_2 \Gamma^{MN} F_{MN}^R N_2
\right] \nonumber \\
&\supset& 
c \int_0^{z_m} dz \frac{i}{z^3} 
\left[
\bar{N}_{1L} \Gamma^{\mu z} F_{\mu z}^V N_{1L} + \bar{N}_{1R} \Gamma^{\mu z} F_{\mu z}^V N_{1R} 
- (1 \leftrightarrow 2)
\right] \nonumber \\
&&+ c \int_0^{z_m} dz \frac{i}{z^3} 
\left[
\bar{N}_{1L} \Gamma^{\mu z} F_{\mu z}^A N_{1L} + \bar{N}_{1R} \Gamma^{\mu z} F_{\mu z}^A N_{1R} 
+ (1 \leftrightarrow 2)
\right] \nonumber \\
&\supset& 
-c \int_0^{z_m} dz \frac{1}{z^3} 
\left[
\bar{N}_{1L} \gamma^{\mu} \gamma^5 (\partial_\mu V_z - \partial_z V_\mu) N_{1L} 
+ \bar{N}_{1R} \gamma^{\mu} \gamma^5 (\partial_\mu V_z - \partial_z V_\mu) N_{1R} \right. \nonumber \\
&& \left. - (1 \leftrightarrow 2) \right] \nonumber \\
&& -c \int_0^{z_m} dz \frac{1}{z^3} 
\left[
\bar{N}_{1L} \gamma^{\mu} \gamma^5 (\partial_\mu A_z - \partial_z A_\mu) N_{1L} 
+ \bar{N}_{1R} \gamma^{\mu} \gamma^5 (\partial_\mu A_z - \partial_z A_\mu) N_{1R} \right. \nonumber \\
&& \left. + (1 \leftrightarrow 2) \right] 
\label{anncplg}
\eea
where $c$ is a constant which is determined by the anomalous magnetic dipole moments 
of the proton and neutron as discussed in \cite{HKSY}. 
In the second and third lines, we extract only the $\mu z$ component relevant to the meson-nucleon coupling. 
In the last expression, the commutator in the field strength is dropped 
and the relation between 4D and 5D gamma matrices $\gamma^5 = -i \Gamma^z$ is used. 

Thus, we obtain the vector and axial-vector meson-nucleon couplings as follows:
\bea
g_{v^n N^l N^l} &\equiv& \int_0^{z_m} dz \frac{1}{z^4} \left[ f_n^V + c z \partial_z f_n^V \right] 
\left[ |f_{1L}^l|^2 + |f_{1R}^l|^2 \right], \\
g_{a^n N^l N^l} &\equiv& \int_0^{z_m} dz \frac{1}{z^4} \left[ f_n^A + c z \partial_z f_n^A \right] 
\left[
|f_{1R}^l|^2 + |f_{1L}^l|^2 
\right]  
\label{ann}
\eea
where $n,l$ are the KK mode indices of (axial) vector meson and nucleons. 

As a bonus, the derivative pion-nucleon coupling $\partial_\mu \pi NN$  is also included in the Pauli term, 
\bea
g_{\partial \pi N^l N^l} \equiv -c \int_0^{z_m} dz \frac{1}{z^3} f^\pi 
\left[
|f_{1L}^l|^2 + |f_{1R}^l|^2 
\right]. 
\eea
This coupling constant is related to $g_{\pi N N}$ through the Goldberger-Treiman relation
$g_{\partial \pi N N} = m_\pi/(2m_N) g_{\pi N N}$, and the consistency can be checked. 

We summarize the numerical results for the above couplings in Tables \ref{results1} and \ref{results2}.  
\begin{table}
\begin{center}
\begin{tabular}{|c|c|c|c|}
\hline
$z_m^{-1}$~(GeV) & $g_{\rho NN}$ & $g_{a_1 NN}$ & $g_{\pi NN}$ \\
\hline
0.205 & 0.08 & 0.42 & -1320 \\
0.33 & -0.54 & -4.09 & -25.6 \\
0.4 & 0.21 & 1.87 & -14.9 \\
0.5 & 0.24 & 3.04 & -13.5 \\
0.6 & 0.32 & 4.16 & -13.2 \\
0.7 & 0.41 & 4.65 & -13.4 \\
0.8 & 0.42 & 3.92 & -13.9 \\
0.9 & 0.23 & 1.45 & -14.9 \\
1.0 & -0.33 & -2.93 & -16.4 \\
\hline
\end{tabular}
\caption{Table of numerical results for various meson-nucleon couplings. 
}
\label{results1}
\end{center}
\end{table}
As an illustration, we have calculated the rho meson-nucleon coupling, 
the $a_1$ meson-nucleon coupling and the pion-nucleon coupling (which has already been calculated in \cite{HIY}). 
In this calculation, we regard the infrared cutoff scale $z_m^{-1}$ 
and Yukawa coupling constant for nucleon $g_Y$ as free parameters. 
We have chosen various input parameters (bare quark mass, chiral condensate and 5 dimensional gauge coupling) as 
\bea
m_q = 2.34~{\rm MeV}, \quad \sigma^{1/3} = 311~{\rm MeV}, \quad g_5 = 2 \pi. 
\eea
These values reproduce the pion mass $m_\pi = 140$ MeV \cite{ekss}. 

In Table \ref{results1}, we have fixed Yukawa coupling $g_Y=9.182$ as taken in \cite{HKSY} and 
calculated couplings $g_{\rho NN}, g_{a^1 NN}$ and $g_{\pi NN}$ for various IR cutoff scales $z_m^{-1}$. 
One can see that if the IR scale $z_m^{-1}$ is around 0.7, 
the observed value of the pion-nucleon coupling $g_{\pi NN}=13.6$ is well reproduced. 
The rho meson-nucleon coupling is however at most 10 percent of experimental value $g_{\rho NN} =4.2 \sim 6.5$. 
\begin{table}
\begin{center}
\begin{tabular}{|c|c|c|c|c|}
\hline 
$z_m^{-1}$~(GeV) & $g_Y$ & $g_{\rho NN}$ & $g_{a_1 NN}$ & $g_{\pi NN}$ \\
\hline
0.6 & 26.5 $\sim$ 26.9 & -4.3 $\sim$ -6.2 & -8.2 $\sim$ -10.5 & -20.1 $\sim$ -22.0 \\
0.7 & 33.6 $\sim$ 34.0 & -5.1 $\sim$ -6.2 & -10.1 $\sim$ -11.4 & -19.8 $\sim$ -20.7 \\
0.8 & 38.6 $\sim$ 40.2 & -4.2 $\sim$ -6.4 & -10.0 $\sim$ -13.1 & -18.8 $\sim$ -20.5 \\
0.9 & 42.5 $\sim$ 44.1 & -5.1 $\sim$ -6.5 & -13.0 $\sim$ -15.1 & -19.8 $\sim$ -20.9 \\
1.0 & 39 $\sim$ 43.8 & -4.2 $\sim$ -6.5 & -13.7 $\sim$ -17.6 & -19.9 $\sim$ -21.7 \\
\hline
\end{tabular}
\caption{Yukawa coupling dependence for various meson-nucleon couplings with fixed $z_m^{-1}$.}
\label{results2}
\end{center}
\end{table}

In Table \ref{results2}, we have tried to improve the results in Table \ref{results1} 
by fixing the IR scale $z_m^{-1}$, but changing Yukawa coupling $g_Y$.  
We have found a parameter region of the Yukawa coupling $g_Y$ 
where the experimental value of rho meson-nucleon coupling is reproduced.\footnote{For $0.2 \le z_m^{-1} \le 0.6$, 
we could not find a viable parameter region for Yukawa coupling 
where the observed rho meson-nucleon coupling is realized.} 
In this region of parameter space, the pion-nucleon coupling does not deviate from the experimental value so much. 
On the other hand, the $a_1$ axial vector meson-nucleon coupling is significantly changed compared to the pion-nucleon coupling. 

Here is a comment on comparison of our results 
with those analyzed by using skyrmions as baryons \cite{HSS}. 
We found that the both results qualitatively agree, 
namely the coupling between the rho meson and nucleons $g_{\rho NN}$ reproduces very well, 
but the pion-nucleon coupling $g_{\pi NN}$ deviates around 50\% from the experimental data. 
As for the coupling between the axial-vector meson and nucleons $g_{a_1 NN}$, 
we cannot say anything about it since we have no experimental data to compare. 

Although our obtained results are relatively good as a first step, 
we certainly need to improve the results. 
For that aim, we might have to take into account other background geometry, 
quantum gravity and stringy corrections, 
and a possible anomalous dimension to the baryon operator and so on 
beyond our simplified approach.

\section{Summary}

In this paper, motivated by recent developments of gauge/string
duality applied to hadron physics, we have considered the bottom-up
approach called AdS/QCD. Unlike the top-down approach, one cannot
precisely determine the bulk geometry as well as the coupling strength
among the bulk fields. However symmetries in the system, which are
originally global on the boundary and are lifted up to local in the bulk,
can be fully utilized. This is a similar situation to chiral Lagrangian
in QCD at low energies. 

In the later part of this paper, 
we have discussed the conjectured scenario called ``holographic unification" of all the mesons and baryons. 
As a concrete example, 
we have considered the meson-nucleon couplings and found some interesting results.
One can further proceed with this idea to compute other physical quantities 
such as tensor couplings as well as other meson-nucleon couplings including scalar mesons and hyperons.

\vspace{.3cm}

\section*{Note Added}
After completing our paper, we noticed a paper \cite{AHPS} with some overlaps 
in the calculation of the meson-nucleon couplings. 

\section*{Acknowledgments}
The authors would like to thank H. U. Yee for useful comments and informations 
and T. Hatsuda for helpful comments. 
One of the authors (N.M.) was supported in part 
by the Grant-in-Aid for Scientific Research 
of the Ministry of Education, Science and Culture, No.18204024. 


\begin{thebibliography}{99}

\bibitem{Mal}
J. Maldacena, Adv. Theor. Math {\bf 2}, 231 (1998).

\bibitem{review}
O. Aharony, S. Gubser, J. Maldacena, H. Ooguri and Y. Oz, Phys. Rept. {\bf 323}, 183 (2000). 

\bibitem{beta}
S. Mandelstam, Nucl. Phys. {\bf B213}, 149 (1983); 
L. Brink, O. Lindgren and B. E. W. Nilsson, Phys. Lett. {\bf B123}, 323 (1983).



\bibitem{holography}
L. Susskind, J. Math. Phys. {\bf 36}, 6377 (1995); 
G. 'tHooft, hep-th/0003004.

\bibitem{qgp}
As a good textbook, see K. Yagi, T. Hatsuda and Y. Miake, "Quark-gluon plasma", Cambridge 
Univ. Press (2005).


\bibitem{csc}
As a recent review, see M. Alford, A. Schmitt, K. Rajagopal and T. Schafer, 
Rev. Mod. Phys. {\bf 80}, 1455 (2008).

\bibitem{wilson}
J. Maldacena, Phys. Rev. Lett. {\bf 80}, 4859 (1998).

\bibitem{glueball}
C. Csaki, H. Ooguri, Y. Oz and J. Terning, JHEP {\bf 9901}, 017 (1999).

\bibitem{karch-katz}
A. Karch and E. Katz, JHEP {\bf 0206}, 043 (2002).

\bibitem{myers}
M. Kruczenski, D. Mateos, R. Myers and D. J. Winter, JHEP {\bf 0307}, 049 (2003); 
J. Babington, J. Erdmenger, N. Evans, Z. Guralnik and I. Kirsch, Phys. Rev. {\bf D69},
066007 (2004). For a recent review, J. Erdmenger, N. Evans, I. Kirsch and E. Threlfall,
Eur. Phys.J. {\bf A35}, 81 (2008).

\bibitem{ss}
T. Sakai and S. Sugimoto, Prog. Theor. Phys. {\bf 113}, 843 (2005). ibid {\bf 114}, 1083 (2005).

\bibitem{witten}
E. Witten, Adv. Theor. Math {\bf 2}, 505 (1998)

\bibitem{ss-baryon}
H. Hata, T. Sakai, S. Sugimoto and S. Yamato, hep-th/0701280; 
K. Nawa, H. Suganuma and T. Kojo, Phys. Rev. {\bf D75}, 086003 (2007); 
D. K. Hong, M. Rho, H. U. Yee and P. Yi, Phys. Rev. {\bf D76}, 061901 (2007). 

\bibitem{deconstruction}
D. T. Son and M. Stephanov, Phys. Rev. {\bf D69}, 065020 (2004).

\bibitem{hls}
For a recent review of hidden local symmetry, see M. Harada and K. Yamawaki,
Phys. Rept. {\bf 381}, 1 (2003).

\bibitem{ekss}
J. Erlich, E. Katz, D. T. Son and M. Stephanov, Phys. Rev. Lett. {\bf 95}, 261602 (2005).

\bibitem{drp}
L. Da Rold and A. Pomarol, Nucl. Phys. {\bf B721}, 79 (2005).

\bibitem{gkpw}
S. Gubser, I. R. Klebanov and A. M. Polyakov, Phys. Lett. {\bf B428}, 105 (1998); 
E. Witten, Adv. Theor. Math. Phys. {\bf 2}, 253 (1998).

\bibitem{shifman}
M. Shifman, hep-ph/0507246.

\bibitem{bulk1}
K. Ghoroku, N. Maru, M. Tachibana and M. Yahiro, Phys. Lett. {\bf B633}, 602 (2006); 
A. Karch, E. Katz, D. T. Son and M. Stephanov, Phys. Rev. {\bf D74}, 015005 (2006); 
J. P. Shock and F. Wu, JHEP {\bf 0608}, 023 (2006); 
J. P. Shock, F. Wu, Y. L. Wu and Z. Xie, JHEP {\bf 0703}, 064 (2007); 
T. Gherghetta, J. I. Kapusta and T. M. Kelly, 0902.1998 [hep-ph]. 

\bibitem{bulk2}
J Hirn and V. Sanz, JHEP {\bf 0512}, 030 (2005); Nucl Phys. Proc. Suppl. {\bf 164}, 273 (2007); 
J. Hirn, N. Rius and V. Sanz, Phys. Rev. {\bf D73}, 085005 (2006); 
J. P. Shock and F. Wu, JHEP {\bf 0608}, 023 (2006); 
H. R. Grigoryan and A. V. Radyushkin, Phys. Rev. {\bf D78}, 115008 (2008); 
S. K. Domokos and J. A. Harvey, Phys. Rev. Lett. {\bf 99}, 141602 (2007); 
Y. Kim, J. P. Lee and S. H. Lee, Phys. Rev. {\bf D75}, 114008 (2007); 
H. C. Kim and Y. Kim, JHEP {\bf 0810}, 011 (2008).

\bibitem{scalar}
L. Da Rold and A. Pomarol, JHEP {\bf 0601}, 157 (2006); 
K. Ghoroku, N. Maru, M. Tachibana and M. Yahiro in \cite{bulk1}; 
E. Katz and M. Schwartz, JHEP {\bf 0708}, 077 (2007); 
P. Colangelo, F. De Fazio, F. Giannuzzi, F. Jugeau and S. Nicotri, Phys. Rev. {\bf D78}, 055009
(2008); 
C. Wang, S. He, M. Huang, Q-S. Yan and Y. Yang, 0902.0864 [hep-ph].

\bibitem{tensor}
E. Katz, A. Lewandowski and M. Schwartz, Phys. Rev. {\bf D74}, 086004 (2006).

\bibitem{glueballs}
P. Colangelo, F. De Fazio,  F. Jugeau and S. Nicotri, Phys. Lett. {\bf B652}, 73 (2007).

\bibitem{baryons}
D. K. Hong, T. Inami and H. U. Yee, Phys. Lett. {\bf B646}, 165 (2007); 
A. Pomarol and A. Wulzer, JHEP {\bf 0803}, 051 (2008), Nucl. Phys. {\bf B809}, 347 (2009); 
A. Cherman, T. D. Cohen and M. Nielsen, 0903.2662 [hep-ph]; 
A. Pomarol and A. Wulzer, 0904.2272 [hep-ph].

\bibitem{exotics}
H. C. Kim and Y. Kim, JHEP {\bf 0901}, 034 (2009).

\bibitem{regge}
H. Boschi-Filho, N. R. F. Braga and H. L. Carrion, Phys. Rev. {\bf D73}, 047901 (2006); 
A. Karch, E. Katz, D. T. Son and M. Stephanov in \cite{bulk1}; 
O. Cata, Phys. Rev. {\bf D75}, 106004 (2007); 
H. Forkel, M. Beyer and T. Frederico, JHEP {\bf 0707}, 077 (2007); 
T. Huang and Z. Fuo, Eur. Phys. J. {\bf C56}, 75 (2008); 
M. Huang, Q. S. Yan and Y. Yang, 0710.0998 [hep-ph], Prog. Theor. Phys. Suppl. {\bf 174}, 334 (2008); 
A. Vega and I. Schmidt, 0811.4638 [hep-ph]; 
S. S. Afonin, 0902.3959 [hep-ph], 0903.0322 [hep-ph].

\bibitem{potential}
O.~Andreev, Phys.\ Rev.\  D {\bf 73}, 107901 (2006);
O.~Andreev and V.~I.~Zakharov, Phys.\ Rev.\  D {\bf 74}, 025023 (2006); 
C. D. White, Phys. Lett. {\bf B652}, 79 (2007); 
W. Y. Wen, Int. J. Mod. Phys. {\bf A23}, 4533 (2008); 
M. V. Carlucci, F. Giannuzzi, G. Nardulli, M. Pellicoro and S. Stramaglia, Eur. Phys. J. {\bf C57},
569 (2008).

\bibitem{finite}
K. Ghoroku and M. Yahiro, Phys. Rev. {\bf D73}, 125010 (2006); 
O. Andreev and V.~I. Zakharov, Phys.\ Lett.\  B {\bf 645}, 437 (2007);
C. Herzhog, Phys. Rev. Lett. {\bf 98}, 091601 (2007); 
E. Nakano, S. Teraguchi and W. Y. Wen, Phys. Rev. {\bf D75}, 085016 (2007); 
R. G. Cai, J. P. Shock, JHEP {\bf 0708}, 095 (2007); 
K.-i. Kim, Y. Kim and S. H. Lee, 0806.3114 [hep-ph]; 
Y. Kim, C. H. Lee, H. U. Yee, Phys. Rev. {\bf D77}, 085030 (2008); 
Y. Kim, S. J. Sin, K. H. Jo and H. K. Lee, JHEP {\bf 0811}, 040 (2008); 
Y. Kim and H. K. Lee, Phys. Rev. {\bf D77}, 096011 (2008); 
H. C. Kim and Y. Kim, JHEP {\bf 0810}, 011 (2008); 
M. Fujita, K. Fukushima, T. Misumi and M. Murata, 0903.2316 [hep-ph]; 
U. Gursoy, E. Kiritsis, L. Mazzanti and F. Nitti, 0903.2859 [hep-ph]; 
U. Gursoy, Mod. Phys. Lett. {\bf A23}, 3349 (2009).

\bibitem{ff}
H. R. Grigoryan and A. V. Radyushkin, Phys. Lett. {\bf B650}, 421 (2007), Phys. Rev. {\bf D76}, 095007 (2007); 
H. J. Kwee and R. F. Lebed, JHEP {\bf 0801}, 027 (2008), Phys. Rev. {\bf D77}, 115007 (2008); 
H. R. Grigoryan and A. V. Radyushkin, Phys. Rev. {\bf D76}, 115007 (2007); 
Z. Abidin and C. E. Carlson, Phys. Rev. {\bf D77}, 095007 (2008); 
H. M. Choi and C. R. Ji, Phys. Rev. {\bf D77}, 113004 (2008); 
H. J. Kwee and R. F. Lebed, 0807.4565 [hep-ph]; 
Z. Abidin and C. E. Carlson, Phys. Rev. {\bf D78}, 071502 (2008); 
C. E. Carlson, 0809.4853 [hep-ph]; 
G. Panico and A. Wulzer, 0811.2211 [hep-ph]; 
Z. Abidin and C. E. Carlson, 0903.4818 [hep-ph].

\bibitem{pheno}
T. Hambye, B. Hassanain, J. March-Russell and M. Schvellinger, Phys. Rev. {\bf 74}, 026003 (2006).
ibid {\bf D76}, 125017 (2007); 
A. N. Atmaja and K. Schalm, 0802.1460 [hep-th]; 
A. Cherman, T. Cohen and E. S. Werbos, 0804.1096 [hep-ph]; 
M. Schvellinger, 0806.0568 [hep-th].

\bibitem{other}
N. Evans, J. P. Shock and T. Waterson, Phys. Lett. {\bf B622}, 165 (2005); 
N. Evans and A. Tedder, Phys. Lett. {\bf B642}, 546 (2006); 
N. Evans, A. Tedder and T. Waterson, JHEP {\bf 0701}, 058 (2007); 
C. Csaki and M. Reece, JHEP {\bf 0705}, 062 (2007); 
G. Panico and A. Wulzer, JHEP {\bf 0705}, 060 (2007); 
T. Schafer, Phys. Rev. {\bf D77}, 126010 (2008); 
E. Shuryak, 0711.0004 [hep-ph]; 
B. Batell and T. Gherghetta, Phys. Rev. {\bf D78}, 026002 (2008); 
Y. Kim, P. Ko and X. H. Wu, JHEP {\bf 0806}, 094 (2008); 
T. Cohen, 0805.4813 [hep-ph]; 
J. Erlich, 0812.4976, 0812.5105 [hep-ph]; 
A. Krikun, Phys. Rev. {\bf D77}, 126014 (2008); 
A. Gorsky and A. Krikun, 0902.1832 [hep-ph]; 
F. Jugeau, 0902.3864 [hep-ph].


\bibitem{HIY}
  D.~K.~Hong, T.~Inami and H.~U.~Yee in \cite{baryons}.
 
 
 
 \bibitem{HSS}
   K.~Hashimoto, T.~Sakai and S.~Sugimoto,
  Prog. Theor. Phys. {\bf 120}, 1093 (2008).
  
  \bibitem{HKSY}
  D.~K.~Hong, H.~C.~Kim, S.~Siwach and H.~U.~Yee,
  JHEP {\bf 0711}, 036 (2007).
  
  \bibitem{AHPS}
  H.~C.~Ahn, D.~K.~Hong, C.~Park, and S.~Siwach,
  0904.3731 [hep-ph].
  
 
\end{thebibliography}
\end{document}